# Room Temperature Structural, Magnetic and Dielectric Characteristics of La Doped CuO Bulk Multiferroic

Kumar Brajesh[1], Sudhir Ranjan,[2] Rajeev Gupta[3], Ambesh Dixit[4] and Ashish Garg[1]
[1]Department of Sustainable Energy Engineering
[2]Department of Chemical Engineering
[3]Department of Physics
Indian Institute of Technology Kanpur, Kanpur-208016 (India)
[4]Department of Physics, Indian Institute of Technology Jodhpur, Karwad – 342037 (India)

## Abstract:

In this manuscript, we report room temperature structural, microstructural, optical, dielectric and magnetic properties of CuO and $Cu_{0.995}La_{0.005}O$ ceramics, synthesized by solid-state reaction method. La doping in CuO leads to the evolution of compact and dense microstructure with reduced porosity. Due to noticeable differences in the ionic radii of $Cu^{2+}$ (0.73Å) and $La^{3+}$ (1.03Å), La doping creates vacancy defects which induces considerable strain in the CuO lattice resulting in a reduction in the lattice parameters and cell volume. However, both ceramics processes similar monoclinic structure with C2/c space group. Detailed characterization using XPS, Raman and FTIR spectroscopy confirmed the incorporation of the $La^{3+}$ in CuO lattice. Interestingly, La doping enhances the dielectric constant by more than three times and results in a reduced leakage current. The onset of large dielectric constant is attributed to dense microstructure and strain/distortion in CuO lattice after La doping. Additionally, the band-gap of $Cu_{0.995}La_{0.005}O$ ceramics decreases which is attributed to increased vacancy defect concentration that creates intermediate dopant energy level within bandgap of CuO matrix. Furthermore, improvement in magnetic and dielectric properties is also discussed and correlated with the grain size in La-doped CuO.

## 1. Introduction

Multiferroic materials with coupling between ferromagnetic and ferroelectric order parameters have extensively been studied in recent years. However, there are very multiferroic materials that are of simple chemistry and that can be synthesized with ease. Copper (II) oxide (CuO) is one such material whose single crystals have been shown to exhibit multiferroic behaviour at ~235K in monoclinic crystal phase [1,2]. CuO(II) oxide is a transition metal oxide possessing monoclinic structure with C2/c space group[3].There are many interesting physical characteristics which make it useful for various applications such as high thermal conductivity, high stability and antimicrobial activity, applications in solar cells, magnetic recording medium, gas sensors, and even in lithium batteries as the anode material [4-7]. Bulk CuO exhibits the multiferroic phase in a narrow temperature range (210 K to 230 K). A transition from a paramagnetic-paraelectric phase to incommensurate or asymmetrical antiferromagnetic (AFM) and ferroelectric state are observed near highest Neel temperature (TN2) ~230 K. In the second phase transition, the order of AFM phase transforms to commensurate AFM phase and ferroelectricity disappears at around the Neel temperature (TN1) 210 K[1,8]. The doping of transition-metal ions in CuO lattice has led to interesting ferromagnetic behaviour at or above room temperature[9,10]. Overall, CuO based ceramics are of great interest due to (i) the high dielectric constant and their potential application in the microelectronic industry, (ii) a lead free environmentally friendly, (iii) low cost, (iv) easily synthesis, and commercial availability at large scale[11]. While the material offers tremendous opportunities for exploration of physical mechanisms of multiferroicity and presence of magnetoelectric coupling, multiferroic transition temperature is too low for realizing any practical application of the material.

The possible ways to overcome this problem could be via composition and strain tuning of the material in both bulk and thin film form which can alter the transition temperatures of ferroics. As shown in various material systems, addition of lanthanides to oxide materials can improve their magnetic, catalytic, and luminescence properties. Recently Tanaka et al. showed that the alkaline and alkaline earth oxides act as the sintering aid to CuO ceramics and these additives at high applied field cause a linearity in their resistivity[12]. Therefore, a high performance ceramics resistor may be developed for the component equipment in generation and delivery power systems, such as transformers, circuit breakers, and arrestors[13]. Dietl et al. showed that transition metal ion doped wide band–gap semiconductor could exhibit room temperature ferromagnetism. In contrast, CuO, a relatively small band-gap semiconductor showed the room temperature magnetic characteristics with transition metal dopants[14]. Hence it would be interesting to examine the characteristics of CuO with rare-earth doping such as dielectric, ferromagnetic, structural, optical, and electrical properties. In this work, we have explored the structural, microstructural, magnetic, dielectric and optical properties of La doped CuO and have conducted various studies to understand the impact of La doping on the properties of CuO.

## 2. Experimental details:

Solid state reaction synthesis was used to prepare polycrystalline CuO and $Cu_{0.995}La_{0.005}O$ powder samples. The powders of high purity lanthanum (III) oxide (99.99%),and copper(II) oxide(99.999%) (all from Sigma Aldrich) were mixed in a stoichiometric ratio using a pestle and mortar. Homogenized powder mixtures were pelletized and calcined at 800°C for 8 hours followed by regrinding, palletization and sintering at 900°C for 10 hours in air. The phase identification of the samples was carried out using PANanalytical's X'Pert PRO X-ray diffractometer. The Nova Nano SEM field-emission scanning electron microscope (FE-SEM) was used for microstructural analysis of the samples. The impedance measurements were carried out using an Agilent 4294A Precision Impedance Analyzer, where an AC signal was applied across the pellets and the output was recorded over the frequency domain (100Hz to 1MHz frequency range). Vibrating sample magnetometer (VSM) was used to measure magnetic properties. X-ray photoelectron spectroscopy (XPS) measurements were performed to analyse chemical states of the samples using PHI 5000 (Versa Probe II, FEI Inc) and XPSPEAK 4.3 software was used to analyse the spectra. Fourier-transform infrared spectroscopy (FTIR, PerkinElmer) was used to study the function groups of the samples in range of wavenumbers 400-4000 $cm^{-1}$. Raman studies was carried out using Raman spectrometer (WiTec) using laser light of wavelength 532 nm. Cary 7000 spectrophotometer (Agilent technologies) was used to record reflectance measurements. Rietveld refinement was carried out using FULLPROF software[15].

# 3. Results and Discussion

## 3.1 Structural and Microstructural analysis

We first conducted X–Ray diffraction (XRD) measurements to study the crystal structure and crystallinity of the synthesized samples. **Figure. 1(a)** and **1(b)** show the room temperature XRD patterns of CuO and $Cu_{0.995}La_{0.005}O$ powder samples which do not show presence of any impurity phase in the samples. All the diffraction peaks were indexed to the crystalline monocline structure of CuO with C2/c space group. Noticeablly, the diffraction peak intensity of La-doped CuO increases in comparison to the pristine sample. We refined all the XRD patterns using FullProf software by considering monoclinic structure with C2/c space group as shown in **Figure 2**. The refinement showed that the crystal structure of CuO powder does not alter after doping $La^{+3}$ ions in the CuO lattice as also evident from the absence of any extra peak in the sample. We did not notice any cubic structured $Cu_2O$ in the XRD patterns. The good-quality of fitting of (002) and ($\bar{2}02$) crystal planes for both the samples, as shown in the inset of **Figure 2(a)** and **2(b)**, indicates the phase matching of the powder XRD patterns. The crystallographical parameters such as lattice parameters, cell volume, atomic positions, goodness-of-fit, are listed in **Table 1** and **2**. **Table 1** shows that both lattice parameters (a, b and c) and the cell volume marginally decreases in La-doped CuO compared to undoped CuO samples which is rather surprising given that the ionic radius of $La^{3+}$ (1.03 Å) is larger than that of $Cu^{2+}$ (0.73 Å). Addition of $La^{3+}$ ions in the host lattice of CuO forms La-Cu-O solid solution due to larger ionic radius of $La^{3+}$ (1.03 Å) than that of $Cu^{2+}$ (0.73 Å) which creates vacancy defects in the CuO lattice resulting in the reduced cell parameters and ultimately contraction of the cell volume. Similar observations have been reported by the *Devi et al*. where La doping in CuO resulted in the contraction of cell parameters due to creation of vacancy defects[16]. Although the change in lattice parameter is very small (at third decimal place), however β increased by $0.02^o$ upon La incorporation, which suggests that La doping in CuO lattice induces lattice distortion which has been further confirmed by Raman analysis, as shown in the subsequent section. Further,we examined the microstructural changes upon La doping which may influence the physical properties of the ceramics. **Figures 3 (a-b)** show the FESEM micrographs of CuO and $Cu_{0.995}La_{0.005}O$ ceramic samples. The microstructures of both samples exhibit agglomerated particles with rather non-uniform grain size distribution. However, compratively speaking, La doped CuO samples possess rather well interconnected grain morphology with more compact and dense microstructure. Formation of agglomerated grains in both the samples can be correlated to high temperature during sintering process which results in increased nucleation rate. Change in the morphology of the La doped samples can be attributed to greater ionic radius of $La^{3+}$ and formation of the La-O-Cu complex that slows down the nucleation rate and crystal growth during sintering, thus corroborating well with contraction in cell volume from XRD results which ultimately leads to compact and dense microstructure[17].

**3.2 Raman and UV-vis spectroscopy analysis:**

Raman scattering in CuO has been reported by Rosen et al and Irwin et al where they found the three peaks centered at 300, 346 and 630 $cm^{-1}$ respectively[18-20]. CuO samples exhibited Raman peaks at ~293.41, 341.81 and 629.03 $cm^{-1}$ as shown in **Figure 4a**. It is worth mentioning that, these wavenumbers are lower than the above reported literature due to size effect . The copper (II) oxide belongs to the $C^6_{2h}$ (C 2/c) space group with two molecules per primitive cell and there are nine-zone centre optical phonon modes with symmetries $4A_u + 5B_u + A_g + 2B_g$; only three $A_g + 2B_g$ mode are Raman active. Peak located at ~293 $cm^{-1}$ is attributed to $A_g$ mode while the peaks located at ~341 and ~630 $cm^{-1}$ could be associated with $B_g$ modes. In the present work, we have shown only active Raman mode for both the samples. Interestingly, as shown in **Figure 4b**. all three Raman active mode peaks in La doped CuO samples red shifted to 290.39 , 336.62, and 627.03 $cm^{-1}$, respectively while the corresponding line width increases as evident from increased full width half maximum (FWHM) values of the $\sim 17.62^0$, $\sim 37.86^0$ and $\sim 24.33^0$ $cm^{-1}$ compared to $13.27^0$, $\sim 16.34^0$ and $\sim 17.43^0$ $cm^{-1}$, respectively for pristine sample corresponding to the three characteristic vibrational modes indicating the phonon confinement effect[20]. This broadening and blue shift of wavenumber confirms the substitution of $La^{3+}$ in the CuO lattice, resulting in microscopic structural disorder in the periodic lattice of La doped CuO and leads to compressive strain within CuO lattice. Furthermore, peak blue shifting could be also attributed to the less agglomeration and reduction in the size of La doped CuO particles which could be associated Heisenberg uncertainty principle indicating relationship between phonon and crystallite size[21,22].

The optical measurements were performed to investigate the optical properties of CuO and La doped CuO ceramics at room temperature by calculating the optical band gaps ($E_{og}$) of both samples as shown in **Figure 5a** and 5**b**. Kubelka-Munk (K-M) model was employed for computing $E_{og}$ of both the samples from the measured reflectance (R) data. The band gap energy of samples was determined by intercept at x-axis in the $[F(R)\ h\nu]^2$ *vs* hν plots, where $F(R)=(1-R)^2/2R$ and hν is photon energy[23]. It is found that $E_{og}$ values of La doped CuO (~1.45eV) is smaller than that of CuO (~1.63eV). Reduction in the band gap of the doped sample can be attributed to the formation to intermediate dopant energy level (due to $La^{3+}$ "4f 5d" localized electronic states) within the bandgap of host CuO matrix upon addition of $La^{3+}$ ions[24]. Moreover, creation of defects and vacancies due to La-doping in CuO matrix also plays an important role in decreasing $E_{og}$ value. Our results are in line with the results reported by Badawi et al[25]. Thus, $E_g$ value of CuO can be tuned by La-doping.

### 3.3 Functional group analysis

**Figure 6** shows the Fourier Infra-Red spectroscopy (FTIR) spectra for analysing the functional groups present in CuO and La doped CuO samples. Vibration bands below 1000 $cm^{-1}$ is attributed to interatomic vibrations from the metal oxide[26]. Characteristic vibration bands located at ~ 546 $cm^{-1}$ and ~ 597 $cm^{-1}$ corresponds to the vibrations arising from metal oxide stretching of monoclinic CuO. Higher frequency vibrational band at ~ 597 $cm^{-1}$ can be ascribed to Cu-O stretching along [-101] direction whereas ~ 546 $cm^{-1}$ band corresponds to the Cu-O stretching occurring along [101] direction.[26] Additionally, presence of $Cu_2O$ phase (vibrational band located at ~ 610 $cm^{-1}$) is also over ruled due to absence of any vibrational band in the range of ~ 605-660 $cm^{-1}$ consistent with XRD results. Interestingly, La doping results in shifting of peak to the higher wave number and becomes broader indicating that La ion is incorporated in the CuO lattice and leads to distortion of the crystal lattice.

### 3.4 Chemical state analysis

XPS measurement was carried out to analyse the chemical states of Cu and La ion in undoped as well as doped samples. **Figure 7**(a-e) shows high-resolution spectra (core XPS spectra) of the Cu 2p, La 3d and O1s, respectively for both the samples. From the **Figure 7**a Cu 2p core level can be deconvoluted into characteristic peaks at 933.55 and 953.65 eV which corresponds to $2p_{3/2}$ and $2p_{1/2}$ for CuO, respectively indicating $Cu^{2+}$ oxidation state while satellite peaks are located at 941.50, 943.50 and 961.75 eV assigned to partially filled 3d orbitals. Interestingly all deconvoluted peaks of Cu 2p core level shifted to lower binding energy in La doped samples (**Figure 7**c). Since Pauling electronegativity of La (1.10) is smaller than Cu (1.90), so electron density of $Cu^{2+}$ is enhanced when La is doped into CuO lattice, resulting in decreased binding energy[27,28]. **Figure 7**e illustrates La 3d core levels centered at 834.88 and 851.43 eV assigned to La $3d_{2/5}$ and La $3d_{3/2}$. Moreover, two satellite peaks are also observed neighbouring to core levels which depicts relocation of electrons of O-2p to an unoccupied La 5f, indicating $La^{3+}$ oxidation state[29,30]. **Figure 7**b shows the deconvoluted high resolution XPS peak of O1s for CuO sample at 529.4 eV corresponding to lattice oxygen ($O_L$) and peak at 530.7 eV assigned to surface oxygen vacancies/defects or surface adsorbed oxygen ($O_V$).

### 3.5 Room temperature study of dielectric constant and dielectric loss

Dielectric and electrical studies provide insight into the localized electric charge carries for better understanding of electrical conduction mechanism and electric polarization. The dielectric parameters, such as dielectric constant ($\varepsilon$) and dissipation factor (tan δ) were evaluated in 100Hz to 1MHz frequency range for CuO and La doped CuO ceramics. The frequency dependence plots for both samples are shown in **Fig. 8**. Although **Fig 8**a displays that the real part of dielectric constant ($\varepsilon'$) decreases sharply with the increase in the low frequency region from 100Hz to 8 KHz for both the samples indicating strong frequency dependence in the low frequency region but La doped CuO samples exhibit higher dielectric constant than the pristine samples. In the high frequency region (8kHz-1MHz), $\varepsilon'$ starts falling slightly with increasing frequency for both samples. The dielectric constant value of La doped CuO ceramics drastically increases by 3.4 times from ~5000 (CuO ceramics) to ~17000. The observed high dielectric constant (~17000) for La doped ceramics may be very useful in microelectronics such as capacitor and memory devices.

Similarly, frequency dependent tan δ plots (**Fig 8b)** for both samples also exhibit strong frequency dependence with doped sample showing higher tan δ than pristine in the low frequency region. The enhanced dielectric constant in La doped CuO ceramics is attributed to the internal barrier layer capacitance (IBLC) effect[31]. According to this effect, the dielectric constant of ceramics can be expressed as $\varepsilon' = \varepsilon_{gb} d/t$, where $\varepsilon_{gb}$ is the dielectric constant of boundary layer, d is the grain size and t is the boundary layer thickness. Observed enhancement in the dielectric constant in doped samples can be also explained by the dense and compact microstructure which results in the lower density of grain boundaries which leads to increased dipole moment per unit volume. Thus, the dielectric constant of the ceramics is directly proportional to the grain size. We observed from FESEM analysis that the grain size of La doped CuO ceramics is relatively larger than that of CuO ceramics, and is probably responsible for observed enhanced dielectric constant for La doped CuO sample. The higher value of tan $\delta$ for La doped CuO at low frequency may be due to the presence of defects in the sample, that play an important role in the conductivity[32].

### 3.6 Leakage analysis

We investigated the electrical leakage characteristics of both the ceramic powders for realising its potential for energy storage applications. Figure 9 shows the current density *vs* electric field curve for both the ceramics. It is observed that doped sample exhibits reduction in leakage current density by 2 orders of magnitude compared to CuO samples. Lower leakage current density in doped ceramic can be attributed to well interconnected grain resulting in dense microstructure that provides lesser leaking path. Additionally, reduction in leakage current density can be also explained on the basis of defect chemistry. Substitution of the $Cu^{2+}$ by $La^{3+}$ results in the creation of extra electrons. Since most of the oxides including CuO is oxygen deficient so $La^{3+}$ doping is expected to reduce the oxygen vacancy concentration leading to reduction in leakage current density [33-35].

### 3.7 Studies of Magnetic properties

The room temperature magnetic hysteresis measurements were performed in a vibrating sample magnetometer (VSM) and corresponding curve is shown in **Fig 10** for CuO and La doped CuO samples. The hysteresis loop shows weak ferromagnetism for both samples. The observed coercive field ~ 347.44 Oe for $Cu_{0.995}La_{0.005}O$ is much higher than the coercive field ~ 69.04Oe for CuO ceramic sample. Doped sample shows higher coercive field than undoped sample, which may be due to the larger grain size (and dense microstructure) for La doped CuO sample as the shape anisotropy of grain size is directly proportional to the coercive field[36]. The non-saturated loops may be due to the presence of clusters of undoped CuO ceramic at room temperature and incomplete formation of lanthanides secondary phases, which are responsible for the observed dominant magnetic behaviour. The observed weak ferromagnetism is superimposed by the paramagnetic contribution of the material and we subtracted the high field paramagnetic contribution and the corrected hysteresis data is shown in **Figure 10** together with inset showing the zoomed view near zero fields. The relatively larger value of electronegativity for Cu (1.90) atom than La (1.10) may induce a partial positive charge on the La atom, when the La atom is bonded with the O atom. This will increase the band gap energy together with oxygen vacancy in La doped CuO sample[37].

## 4. Conclusions

CuO and La doped CuO ceramics were synthesized by the traditional solid-state reaction and their phase purity and crystal structure is confirmed using XRD measurements. Reitveld refinement of the XRD pattern reveals contraction of cell volume in La doped sample together with the lattice distortion which was further supported by Raman spectroscopy. Increased defect concentration in La doped CuO samples reduces the band gap of the doped ceramics. Interestingly, relatively large dielectric constant for La doped CuO ceramic sample is observed, resulting in reduced leakage current and thus, posing as be suitable candidate for various energy storage devices and electronic applications. Additionally, the phase pure monoclinic CuO and La doped CuO samples showed weak ferromagnetism even at room temperature.

**Acknowledgments**: Authors thank the Science and Engineering Research Board (SERB), Government of India, for the financial support through grant no EMR/2017/003486.

# List of Tables

### Table 1. Refined monoclinic (C 2/c) structure parameters of CuO ceramics

| Atoms | x | y | z | B (Å) |
|---|---|---|---|---|
| Cu/La | 0.25 | 0.25 | 0 | 1.67(1) |
| O | 0 | 0.417(4) | 0.25 | 1.40(4) |
| | | | | |
| a =4.684(1), b = 3.426 (1) and c = 5.132(1), β = 99.439(7), Vol. = 81.30(4)$Å^3$, $R_p$ = 3.5%, $R_{wp}$ = 4.79%, $R_{exp}$ = 2.35%,and $\chi^2$=2.35 | | | | |

### Table 2. Refined monoclinic (C 2/c) structure parameters of $Cu_{0.995}La_{0.005}O$ ceramics

| Atoms | x | y | z | B (Å) |
|---|---|---|---|---|
| Cu/La | 0.25 | 0.25 | 0 | 1.82(1) |
| O | 0 | 0.420(4) | 0.25 | 1.55(4) |
| | | | | |
| a =4.682(1), b = 3.423 (8) and c = 5.131(1), β = 99.457(7), Vol. = 81.20(4)$Å^3$, $R_p$ = 3.66%, $R_{wp}$ = 5.02%, $R_{exp}$ = 1.86%,and $\chi^2$=2.54 | | | | |

# List of Figures

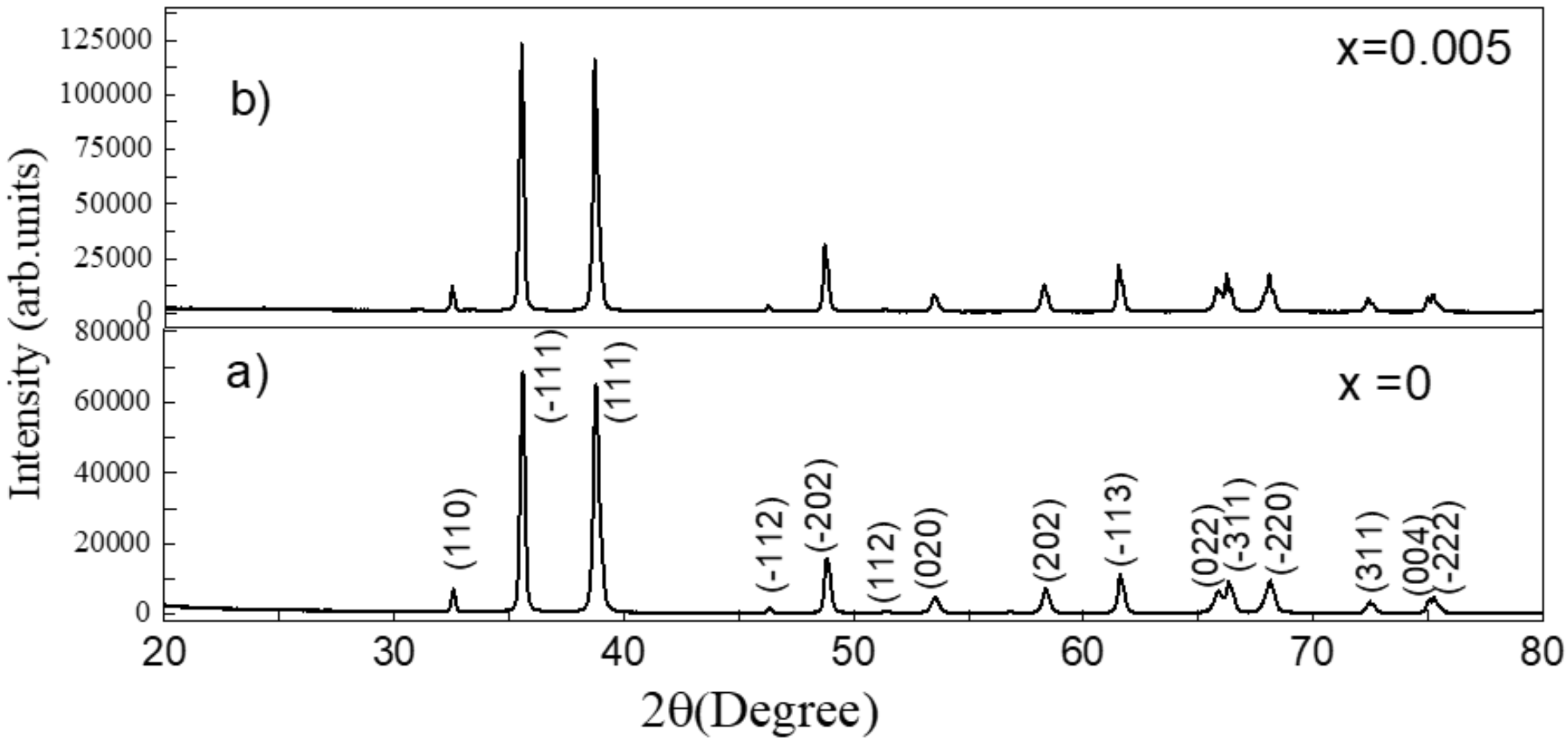

**Figure 1.** X-ray diffraction patterns of (a) La doped CuO and (b)CuO ceramics at room temperature.

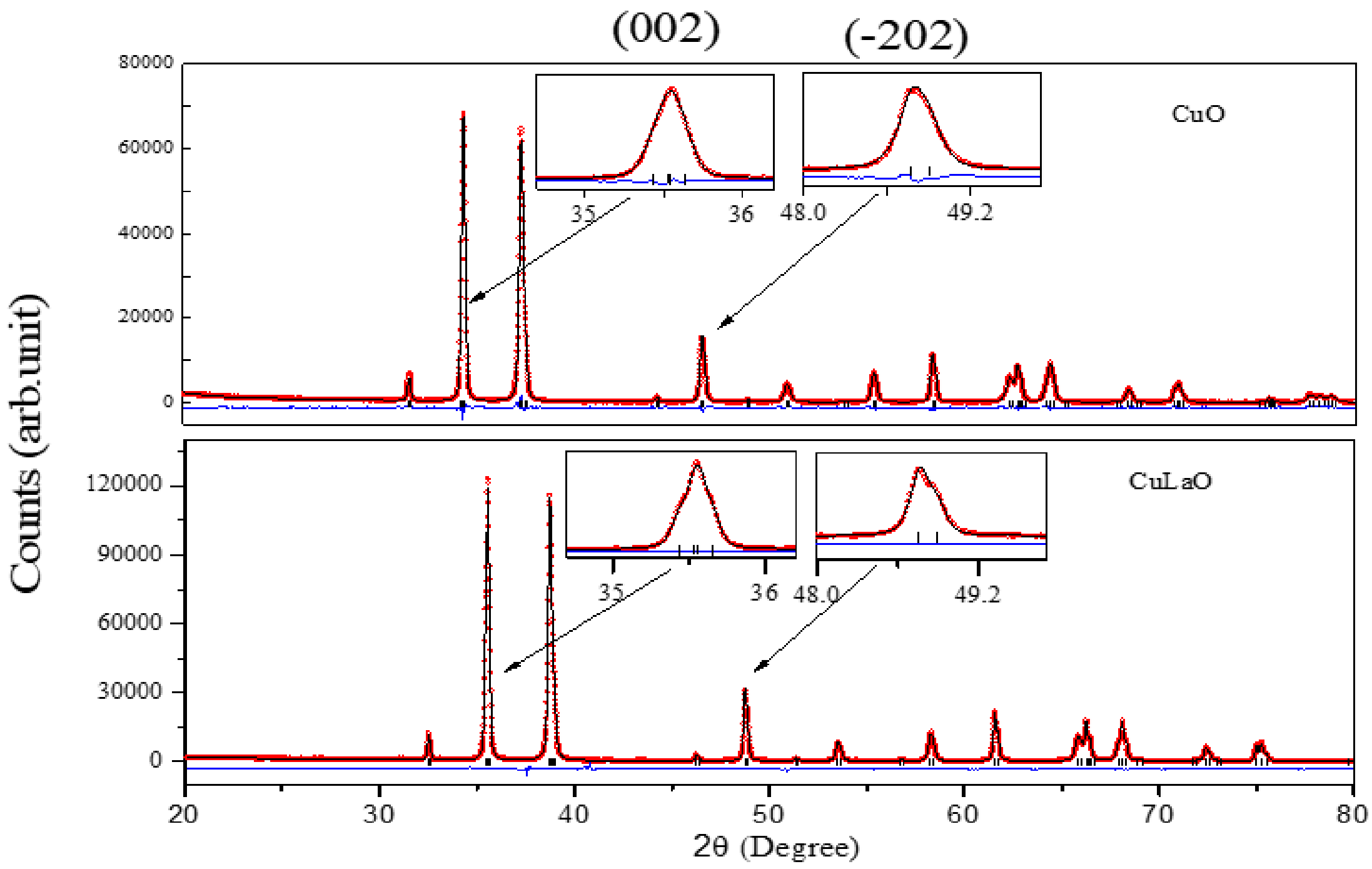

**Figure 2.** Reitveld refinement of X-ray diffraction patterns for CuO and $Cu_{0.9995}La_{0.005}O$ ceramics at room temperature.

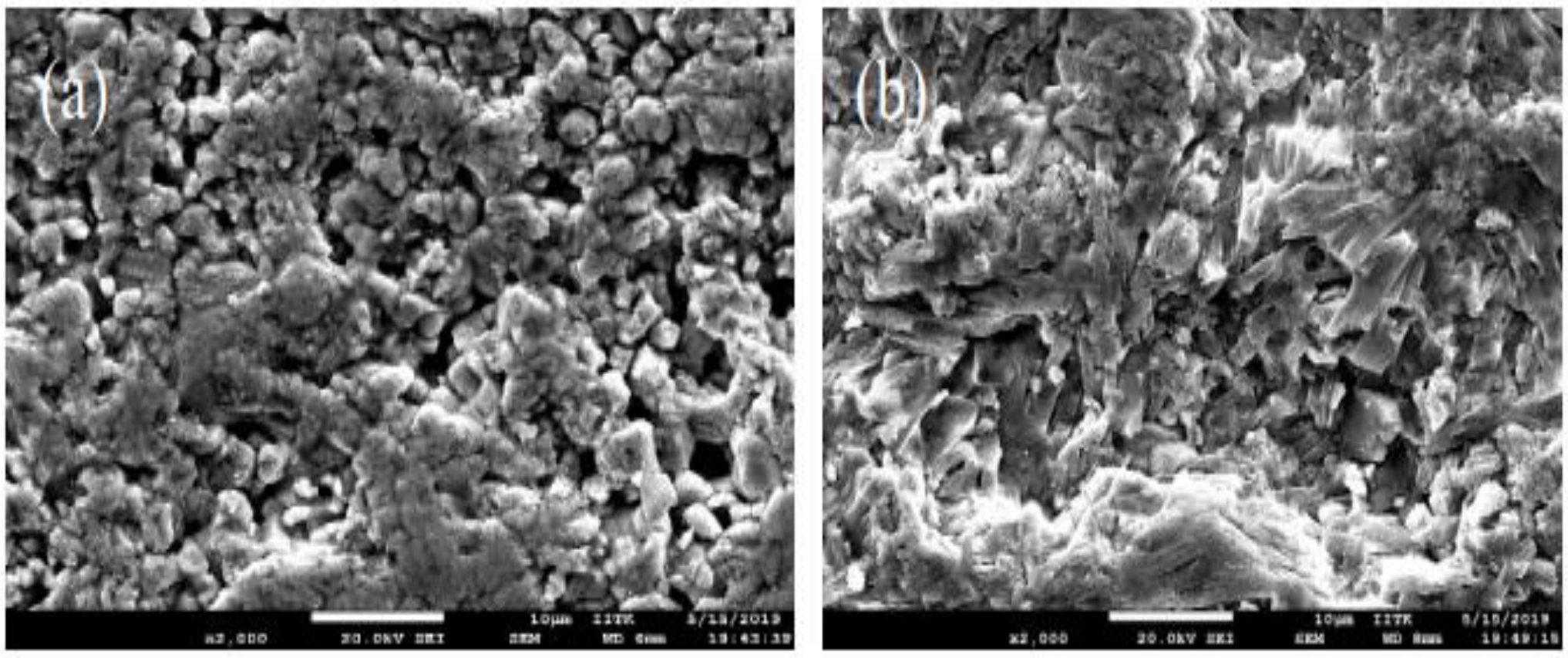

**Figure 3.** SEM micrographs of (a) CuO and (b) $Cu_{0.9995}La_{0.005}O$ ceramics at room temperature.

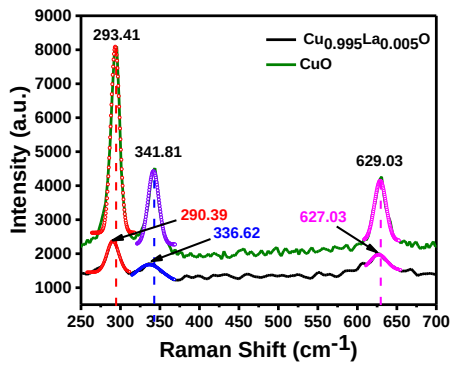

**Figure 4.** Raman spectra with Gaussian fitting of (a) CuO and (b) $Cu_{0.9995}La_{0.005}O$ ceramics at room temperature.

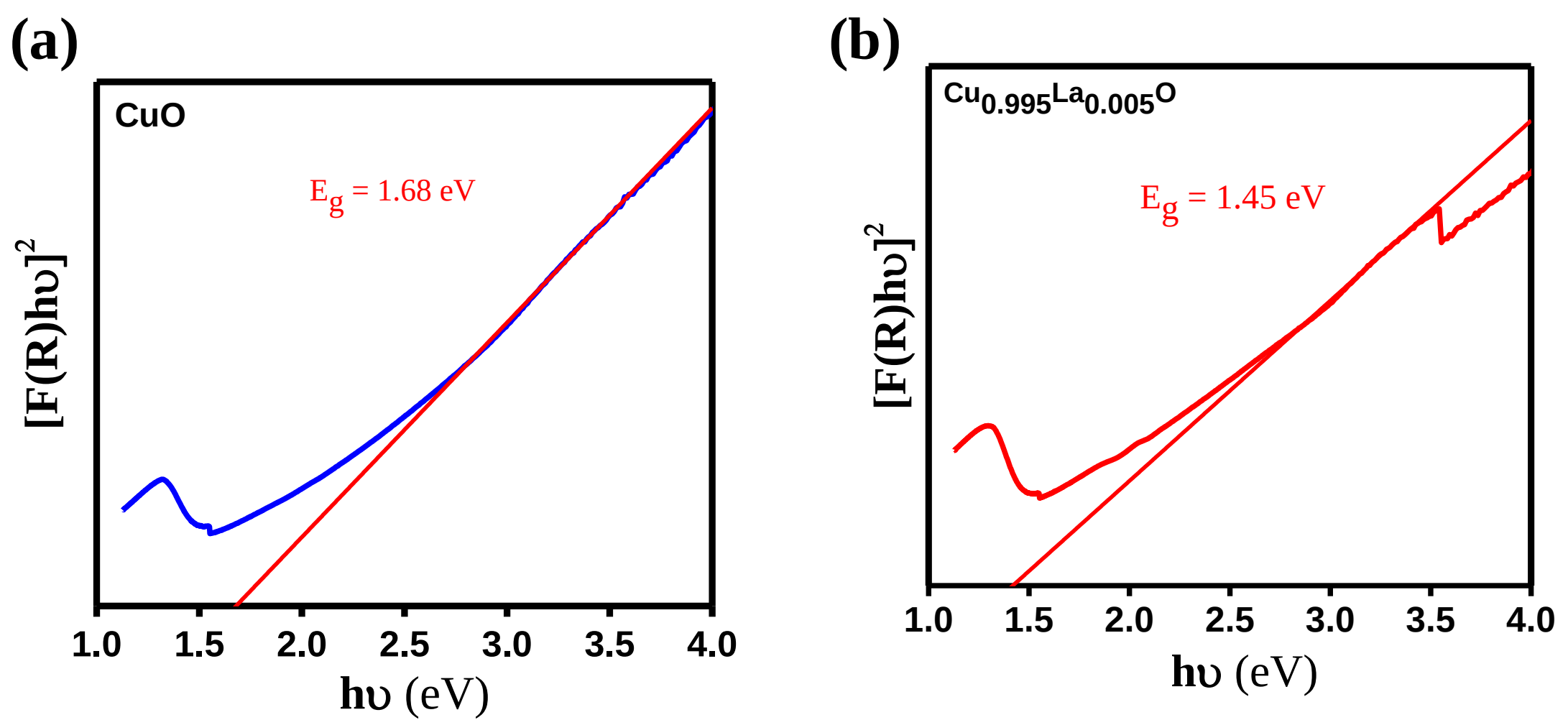

**Figure 5.** $[F(R)\ h\nu]^2$ versus hν plots of (a) CuO and (b) La doped CuO ceramics at room temperature.

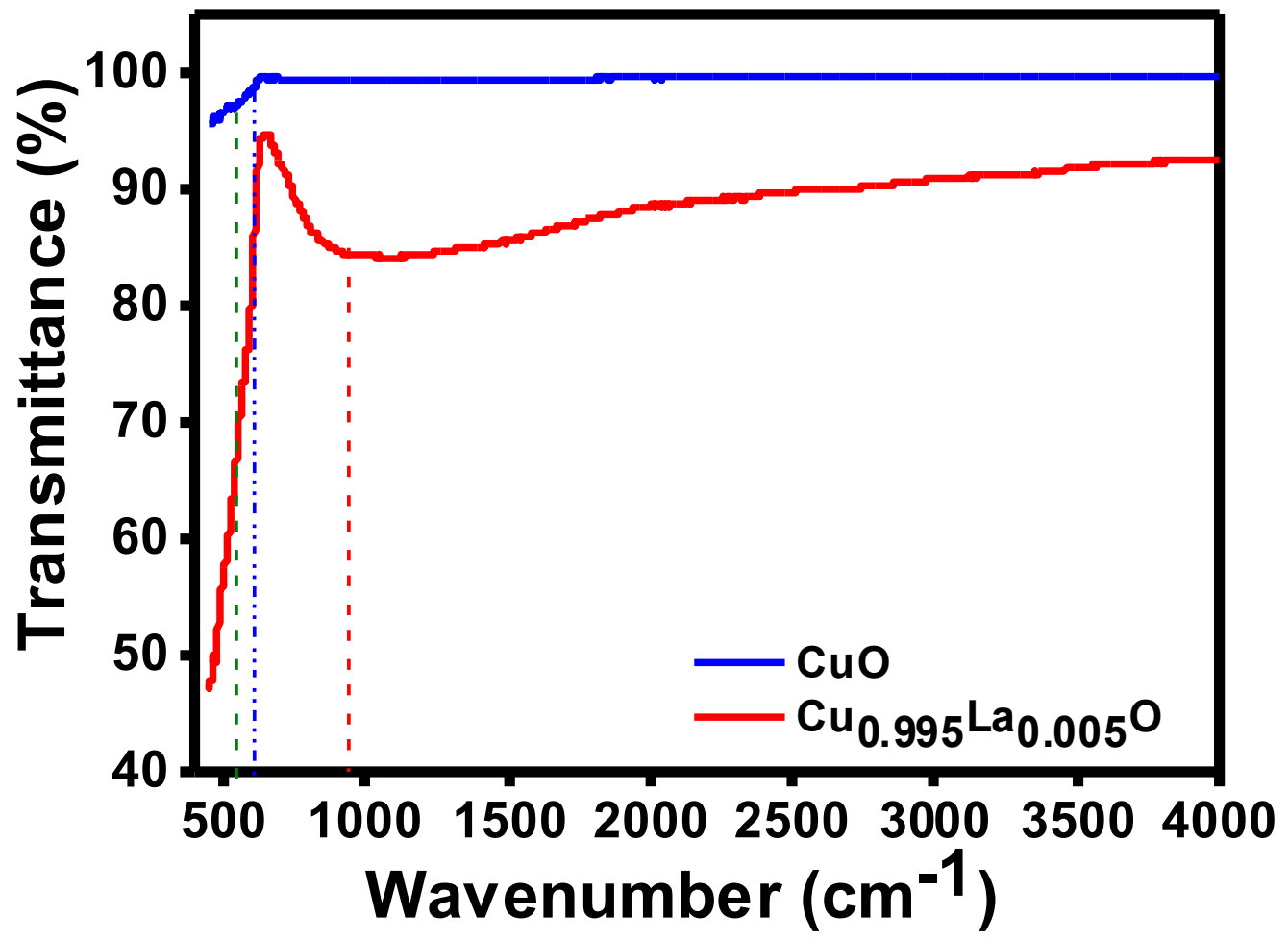

**Figure 6.** FTIR spectra of (a) CuO and (b) La doped CuO ceramics at room temperature.

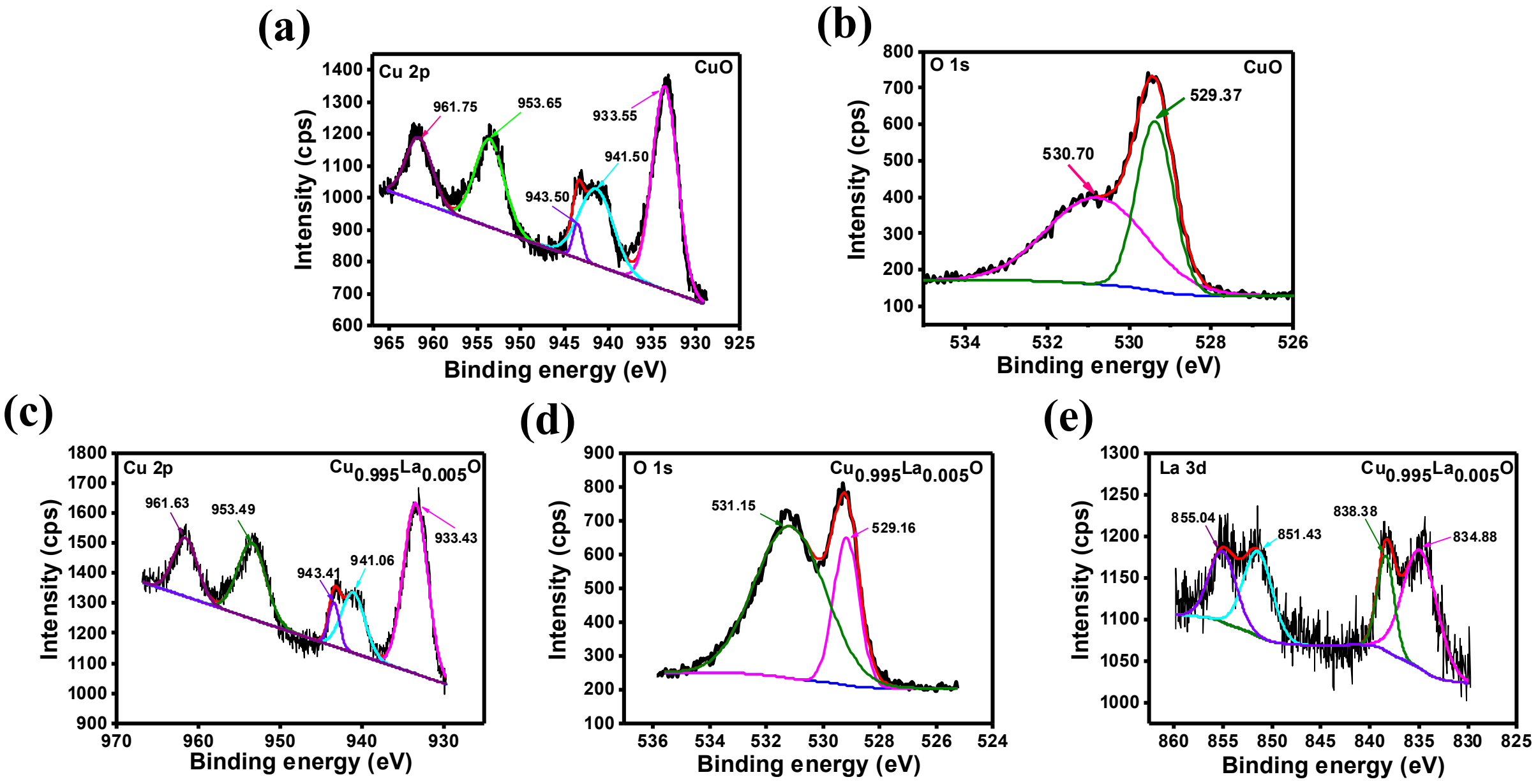

**Figure 7.** High resolution XPS spectra of (a) Cu 2p and (b) O 1s for CuO. High resolution XPS spectra of (c) Cu 2p (d) O 1s and (e) La 3d for $Cu_{0.9995}La_{0.005}O$ ceramic.

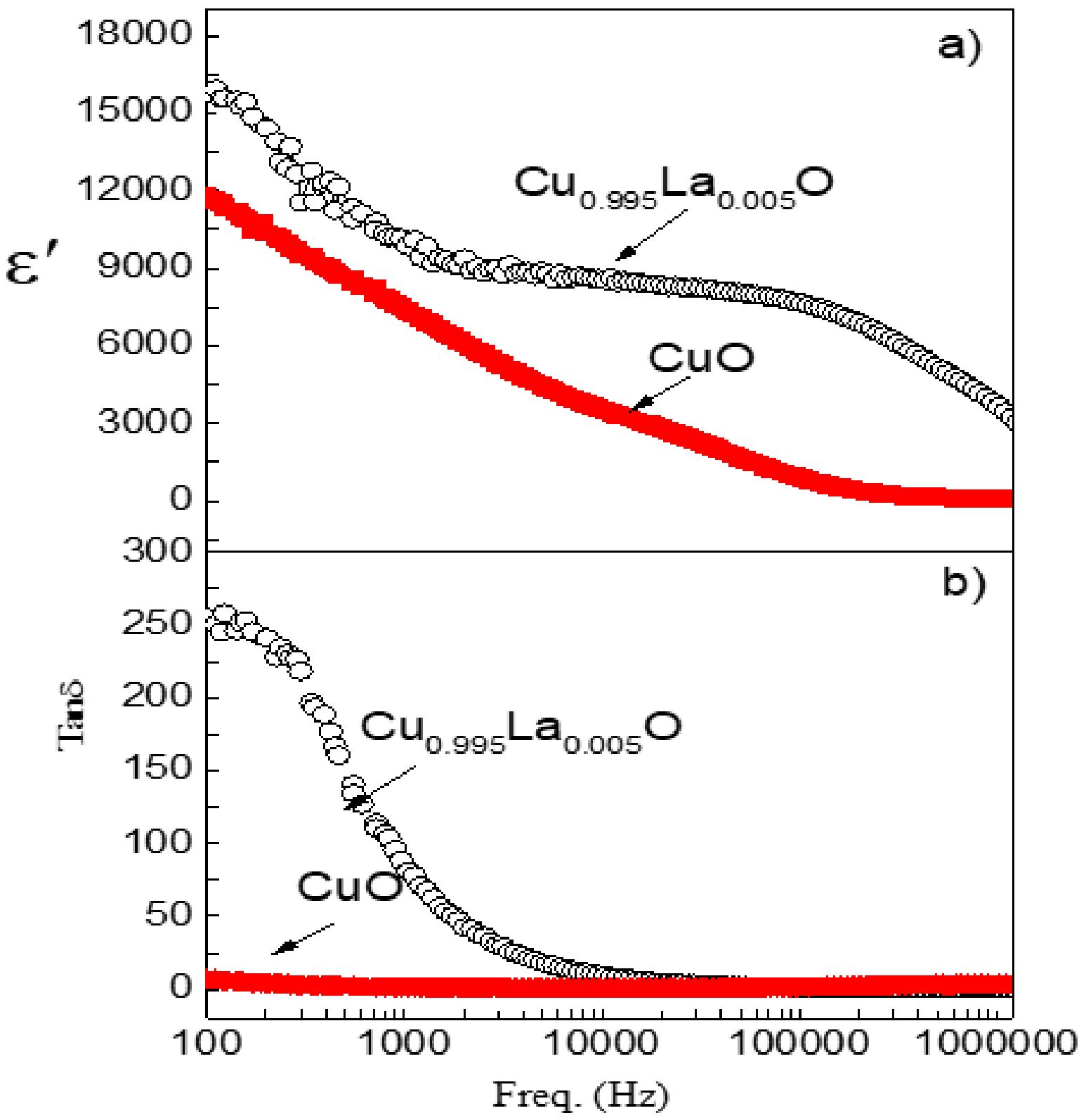

**Figure 8.** The frequency dependent of (a) Dielectric constant of doped and undoped CuO and (b) dissipation factor (tanδ) of doped and undoped CuO at room temperature.

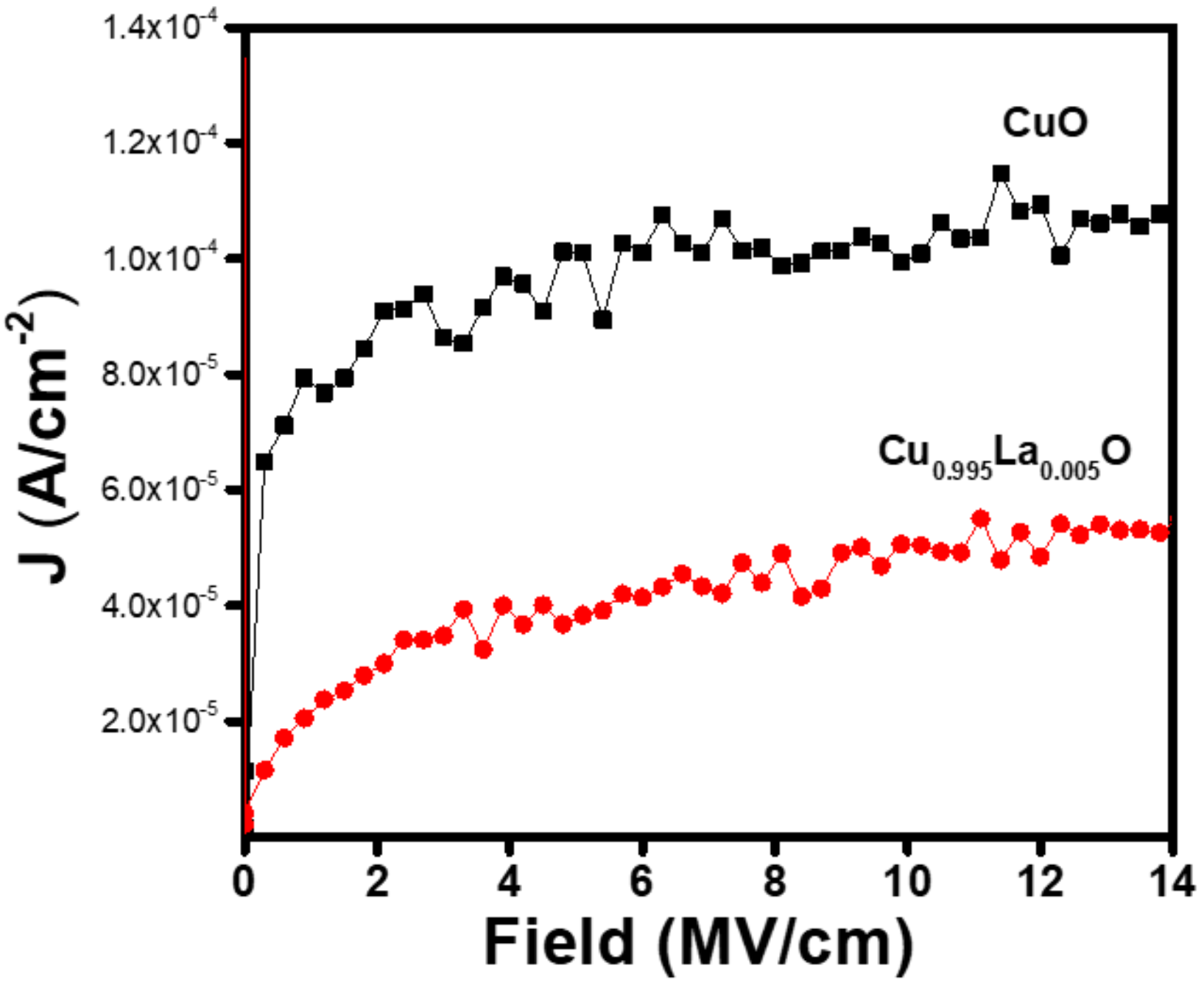

**Figure 9.** Leakage current density plot of CuO and $Cu_{0.9995}La_{0.005}O$ ceramics as a function of electric field.

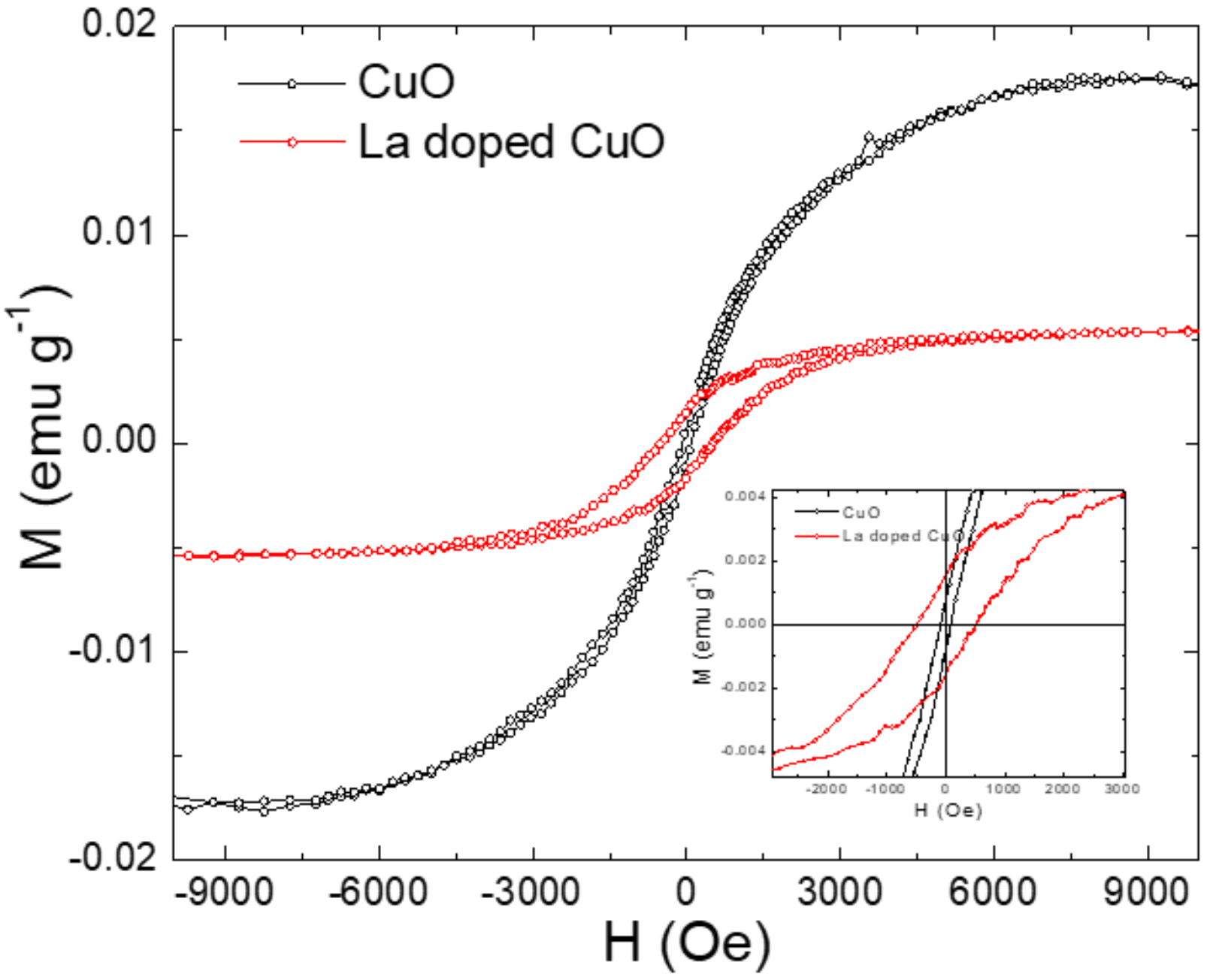

**Figure 10**. Field dependent magnetic moment of (a) CuO and (b) La doped CuO ceramics at room temperature.